\documentstyle[12pt]{article}
\begin{document}
\baselineskip=24pt
\setcounter{page}{1}
\parskip=0pt plus2pt
\textheight=22cm
\begin{titlepage}
\begin{center}
{\Large\bf
Heterogeneities in Supercooled liquids: \\
A Density Functional Study}
\vspace*{.5cm}
\end{center}

\begin{center}
{\it  Charanbir Kaur and Shankar P. Das }\\
{\it
School of Physical Sciences,\\
Jawaharlal Nehru University,\\
New Delhi 110067, India. }

\end{center}

\vspace*{1cm}
\begin{center}
{\Large Abstract}
\end{center}

A metastable state, characterized
by a low degree of mass localization is identified using Density
Functional Theory.
This free energy minimum, located through the proper evaluation of the 
competing terms in the free energy functional, is independent
of the specific form of the DFT used.
Computer simulation results on particle motion indicate 
that this heterogeneous state corresponds to the supercooled state.
\noindent

\vspace*{2cm}
\noindent 
PACS number(s) : 64.70Pf, 64.60 Cn, 64.70 Dv
\end{titlepage}

The properties of supercooled liquids are of much current
interest and are being studied from different approaches
related to their thermodynamic as well as dynamic aspects 
\cite{FV}-\cite{VS}.
A normal liquid is   characterized by a homogenous density and
when it is supercooled below the freezing 
transition it continues to stay in the amorphous phase.
With the increase in density a solid like phase is formed with
the particles being localized around their mean positions on a random
structure. The underlying lattice on which such localized 
motion takes place is related to the time scales of relaxation
 in the supercooled liquid.
While the supercooled liquid starts to attain solid like properties,
 structurally it does not have any long range order like the
 one present in a crystal.
The Heterogeneity in glassy systems over length and time scales
has been studied in several recent works, \cite{UK},
 related to computer simulations.
  Here we consider the heterogeneous density profile in the liquid and
 investigate the question of having  metastable states 
 in between the homogeneous liquid state and the regular 
 crystalline state.
The stability of such a structure has been studied in this work
from a thermodynamic point of view,
using the standard methods of Density Functional
Theory (DFT).
It provides the means to test if a given structure i.e. a configuration
of atoms is the ``most'' stable at a specified temperature and density.
This method has generally been used \cite{RY}-\cite{BC2} for the study
of a liquid  freezing into an ordered crystalline state through
a first order transition.

In practical calculations of the DFT an explicit functional form for
the inhomogeneous density function against which the Free energy
functional is tested, is needed.  One very successful
prescription of density distribution is
 as a superposition of density profiles centered on a Lattice,
\begin{equation}
\label{dens}
\rho (\vec r)=\sum_i (\frac{\alpha}{\pi})^\frac{3}{2}e^{-\alpha {|\vec{r}-\vec{R_i}|}^2} .
\end{equation}
\noindent
where the $\{ {\vec R_i} \} $ denotes the underlying lattice. 
Here $\alpha$ is the variational parameter that characterizes
the width of the
peak, which represents the degree of localization of mass in the system.
Thus the homogeneous liquid state is characterized
by Gaussian profiles of very large width such that 
each provides the same contribution in the sum
at all spatial positions. 
Singh $ et.al. $ \cite{SW} have shown that the distribution
of ${\vec{R_i}} $ over a random lattice determined from the Bernal
packing \cite{BER} allows a metastable minimum for the 
free energy functional for packing  fraction $\eta$ beyond .59.
This is obtained for very large  $\alpha$, that
corresponds to an inhomogeneous and {\em highly localized} density
distribution that has been termed as the ``hard sphere glass''.
This metastable state however is not compatible
with the experimental findings \cite{ST,VS} which find the metastable
supercooled state with a much lower degree of localization.

We have studied this through the various \cite{RY,CA}
formulations of the DFT  for the  density profiles centered
around a random lattice that corresponds to a heterogeneous density 
distribution.
Here we consider the  density profiles that are {\em less} 
localized than the so-called hard sphere glass by 
evaluating the Free energy of the system for  $\alpha$
values that are {\em considerably smaller} than what was previously
 studied.
We also consider the effects of fluctuations in the width around a 
 mean value.
The key result of this study
is the observation of a free energy minimum where the density function
corresponds to the small $\alpha$
region. This minimum is seen apart from the usual high $\alpha$ minimum,
as reported in the earlier
works, \cite{SW,BC,HL} and this state can be reached in a 
{\em continuous } path from both the liquid and the high $\alpha$ state 
on the free energy landscape.

The expression for the total free energy contains two parts -
 the ideal gas term and the interaction term,
$ F[\rho]=F_{id}[\rho] + F_{ex}[\rho] $.
The ideal gas term of the free energy functional (in units of
$ \beta^{-1}$) is given by,
\begin{equation}
\label{ide}
 F_{id}[\rho]=\int d \vec r \rho(\vec r) ( ln [ \wedge ^3 \rho(\vec r)
] -1 )
\end{equation}
\noindent $\wedge$ being the thermal wavelength.
 In the earlier works,  \cite{SW,BC,HL}, where highly localized
 structures have been investigated, generally $\alpha$ was chosen
 to be large ( greater than $ \approx 50 $) and eqn.(\ref{ide}) was 
 approximated by its asymptotic value for large $ \alpha$,
\begin{equation} 
\label{app}
F_{id}[\rho] \approx N [ -\frac{5}{2} + ln(\wedge ^3
(\frac{\alpha}{\pi})^\frac{3}{2} ] 
\end{equation}
However, in the low $ \alpha $ range
where overlapping Gaussians from different sites contribute, 
we evaluate this term exactly from the computation of the integral 
given in (\ref{ide}) as,
\begin{equation}
\label{EX}
{f}_{id}[\rho]=
\int d\vec r \phi(\vec r) \left[ ln \left( \wedge ^3 
\int d \vec R  \phi( \vec r - \vec R) \left( \delta (\vec R) + \rho_o  
g( \vec R) \right) \right) -1 \right]
\end{equation}
\noindent where $ g(\vec R) $ is the site-site correlation function which
provides the structural description 
of the random structure used. We have used 
the Bernal's random structure \cite{BER} generated through the
Bennett's algorithm \cite{B}.
We approximate the $ g(\vec R) $ as,
\begin{equation}
\label{eta0}
g(\vec R) = g_{B}[ R(\frac{\eta}{\eta_o})^{\frac{1}{3}}]
\end{equation}
\noindent where $ \eta $ denotes the average packing fraction and
$ \eta_o $ is used as
a scaling parameter for the structure, \cite{BC,HL} such that at
$ \eta =\eta_o $
Bernal's structure is obtained.
The ideal free energy value evaluated using this exact treatment,
eqn.(\ref {EX}), is shown in Figure 1 for $ \rho_o=1.0 $.
This agrees on extrapolation to the limit
of $ \alpha \rightarrow {0} $  (i.e. $ \rho(\vec r)
\rightarrow {\rho_o})$ result $ i.e. -1$. The exact evaluation starts
approaching the asymptotic ( large $\alpha$) result (dashed line), 
for $ \alpha  > 20  $ within  5\% as shown in the figure.
The interaction part is evaluated using the standard formalism
used by Singh et. al. \cite{SW} with the expression for the
Ramakrishnan-Yussouff (RY) functional,
\begin{equation}
\label{taylor}
{\Delta F}_{ex} = -\frac{1}{2} \int d \vec r_1 \int d \vec r_2
c(|\vec r_1 - \vec r_2|;\rho_o)~ \delta \rho(\vec r_1)~ \delta \rho(\vec r_2)
\end{equation}
that gives the difference in the free energies of the solid and liquid phase
of average density $ \rho_o $. Here $\delta \rho(\vec r) $ is the density
fluctuation from the average value $\rho_o$.
We use the solution of the Percus Yevick equation with Verlet Weiss
correction for the direct correlation function, c(r) \cite{VW,HG}. 
The low $\alpha$
minimum becomes a metastable state between the homogeneous liquid
state $( \alpha \rightarrow 0 )$ and the crystal state beyond
$ \eta = 0.576$ ($\rho_o=1.10$).
In order to clarify this point we have shown in Figure 2, the minimum value
for the difference of the free energy per particle ( for the corresponding $\alpha$ value)
against the density.
The Free energy  corresponding to the ``hard sphere glass'' state 
become metastable w.r.t homogeneous liquid state at average density $ \rho_o = 1.14$
\cite{SW}. Moreover, the new  minimum found in the strongly 
overlapping region (low $\alpha$) can be reached continuously from both
the liquid state minima and the hard sphere glass state.
We also like to stress here that the expansion (\ref{taylor}) for
the Free Energy of the liquid  used in the RY approach is a better 
approximation  for the minima observed at low $\alpha$ than
in the case of the highly localized structure called the hard sphere glass.
In figure 3, we show the free energy evaluated with the density function
obtained for alpha extending to small values.
The  minimum appears at $\alpha \approx 18 $ for  $\rho_o$ 1.12.
The free energy minimum identified in Ref. \cite{SW}  also occurs  but
for very high value of $\alpha$ and
corresponds to highly localized structures referred to as the hard sphere
glass. The Free energy  corresponding to the low $\alpha$ minima
is less than that for the ``hard sphere glass'' state over
the density range considered here.
However, if the back ground lattice is taken as a {\em regular} crystalline
one then the free energy {\em does not} show any
minimum in the  small $\alpha$ region unlike the case
 of an underlying amorphous structure.
This indicates an inhomogeneous structure of strongly overlapping
Gaussians centered around regular lattice sites is ruled out. Indeed
such uniform structures are never seen in simulations. 
This {\em minimum is only seen for the amorphous structure}
which signifies a heterogeneous density distribution. This
can be given a more quantitative form in the following way :
The $ \alpha $ corresponding to the minimum free energy value is inversely
proportional to the root mean square displacement of the particles from
their sites, which also defines the Lindemann \cite{LD}  ratio.
The two minima with the random structure for low and high value
of $\alpha$, respectively found in the present work
and in Ref. \cite{SW,BC,HL} correspond to very different degrees of 
localization of the particles.   
The simulation results, \cite{ST,VS}, show that the Lindemann ratio of
supercooled liquid is approximately three times that of 
crystal at freezing(Figure 2 of \cite{ST}).
We have observed a similar relation for the metastable state
corresponding  to the small $ \alpha $ minimum.
This is shown in the Figure 4 where we plot the respective root mean
square displacement ($ d \sim \frac{1}{\sqrt{\alpha}}$) for both the 
supercooled and the crystal case.
This fact strengthens the case for identifying it with the supercooled
state seen in computer simulations as compared to the highly localized 
hard sphere glass which shows a level of localization close to that of the 
corresponding crystal.
The Barrier height between the liquid and the supercooled phase 
grows with the increase in the average density, as  shown by points in 
the Figure 5. 
It follows a power law increase with the height diverging 
at packing fraction $ \approx 0.62 $ and exponent 1.57
as shown by the solid line in the Figure 5. In all these calculations we have
used $\eta_o~= .64$.
The pressure ( P ) and Bulk
modulus ( E ) of the corresponding structure are computed using 
the first and second derivatives of the total free energy per unit 
volume (\cite{HL}). We show in figure 6, results computed using
the present model, for a random structure corresponding to $\eta_o=.68$
in (\ref{eta0}).
The total free energy is calculated
using the Modified Weighted Density Approx
(MWDA) treatment \cite{CA} with a semi-empirical form of c(r) \cite{CBX}. 

We have also considered the fluctuation
of the Gaussian Density profile's width parameter $\alpha$ over different sites in the
random lattice to incorporate higher degree of heterogeneity.
This is modeled by attaching an independent  probability distribution function, $ P(\alpha_i) $, 
{\em to each lattice site}, that governs $ \alpha $ value. 
We chose P to be Gaussian peaked at $\bar{\alpha}$ and spread ( half width )
$\bar{\alpha} r$, to compute the free energy averaged over the $\alpha$ fluctuations
as a function of the parameters $\bar{\alpha}$ and $r$.
These fluctuations in $ \alpha $ bring about an overall increase in the free
energy of the system i.e. the system stability decreases on account of
increased  heterogeneity at the individual unit of mass concentration. 
 This is also self consistent with the choice of using the direct correlation function
 of a hard sphere system with single size.

The motivation of this study has been to evaluate the stability of
heterogeneous density distributions from a purely thermodynamic viewpoint.
The existence of the free energy minimum corresponding to
a density distribution of overlapping Gaussians centered around an 
amorphous lattice  depicts the supercooled state with a  heterogeneous 
density profile. This state can be reached continuously from
the homogeneous liquid state or the so-called hard sphere glass
minimum. As the density increases the energy barrier to come out of 
this minimum grows.
The identification of this inhomogeneous state is 
indeed linked to the proper evaluation of the ideal gas part of the
free energy. In earlier works this was generally approximated by
the asymptotic formula given in equation (\ref{app}) which works well
for large values of $\alpha$ representing highly localized structures.
In the present work with the proper evaluation of the ideal
gas term for the heterogeneous density, role of configurational
packings on the Free energy is taken into account.
From our comparison with the Lindemann parameter values (Fig. 4) it follows
that this heterogeneous state with less degree of mass localization does agree
with the computer simulation results better, as compared to the so called
Hard Sphere Glass observed by Singh et. al. \cite{SW}.
This new minimum does not occur for the structure centered on
an ordered lattice like the f.c.c., strengthens the case
for the heterogeneous glassy phase.

The qualitative nature of the state is different from the hard sphere glass
state that was identified in the earlier works. We have used here the
Bernel packing to define the underlying lattice but these studies can be
extended to different types of random structure, even
taking results from Computer simulation studies. 
We also like to mention that this new minimum shows up for
different forms of the Density functional Theory and with
different direct correlation functions.
This minima is observed with both the treatments of RY functional and the
MWDA \cite{CA} even with the PY c(r) without
any tail. Indeed improvement of these results can be obtained with a better
input for the structure functions using improved techniques \cite{andreas}.
Including higher order correlations namely the three point functions,
in the expression (\ref{taylor}) is also expected
to account for increased cooperativity at high density.

\noindent
\section*{Acknowledgement}
We thank Prof. Y. Singh and Prof. Chandan Dasgupta for their comments.
SPD acknowledges the support from  grant INT9615212 from NSF.
CK acknowledges financial support from the University Grants Commission
(UGC) of India.

\newpage{}
\section*{Figure Captions}

\vspace*{.5cm}
\noindent
Fig 1 :
Ideal gas part of the free energy per particle ( in units of 
$\beta^{-1}$) vs.  $ \alpha $ ( in units of $\sigma^{-2}$ )
for density $ {\rho_o}^* =1.0 $.
The solid line is obtained
from the exact evaluation, i.e. eqn. (\ref{EX}) and the dashed line is the
result from the approximation.

\vspace*{.5cm}
\noindent
Fig 2 :
Free Energy difference per particle ($\Delta f$)
( in units of $ \beta^{-1}  $ ) 
vs. density in the low $\alpha$ regime.

\vspace*{.5cm}
\noindent
Fig 3 :
Difference in Free Energy per particle ($\Delta f$)
 ( in units of $ \beta^{-1} $ ) vs. $ \alpha $ ( in units of 
$\sigma^{-2}$) in the low $\alpha$ regime ($ {\rho_o}^*=1.12 $).
The inset demonstrates the continuity of the curve near the
 maximum.

\vspace*{.5cm}
\noindent
Fig 4 :
Average displacement $d~$ ( in units of $\sigma$ )
 vs. $\rho_o$. The dashed curve depicts the supercooled 
phase and the solid line  is for the fcc crystal.

\vspace*{.5cm}
\noindent
Fig 5: Barrier height $ h $ (in units of $ \beta^{-1} $) between the 
liquid and the supercooled phase vs. the average packing 
fraction $\eta $.
The solid line is a power law fit to this data.

\vspace*{.5cm}
\noindent
Fig 6:
The  Bulk Modulus E(in units of ($\beta \sigma^3)^{-1}$) vs. density for
the amorphous structure. In the inset, pressure P
(in units of ($\beta \sigma^3)^{-1}$)
vs. density is shown.

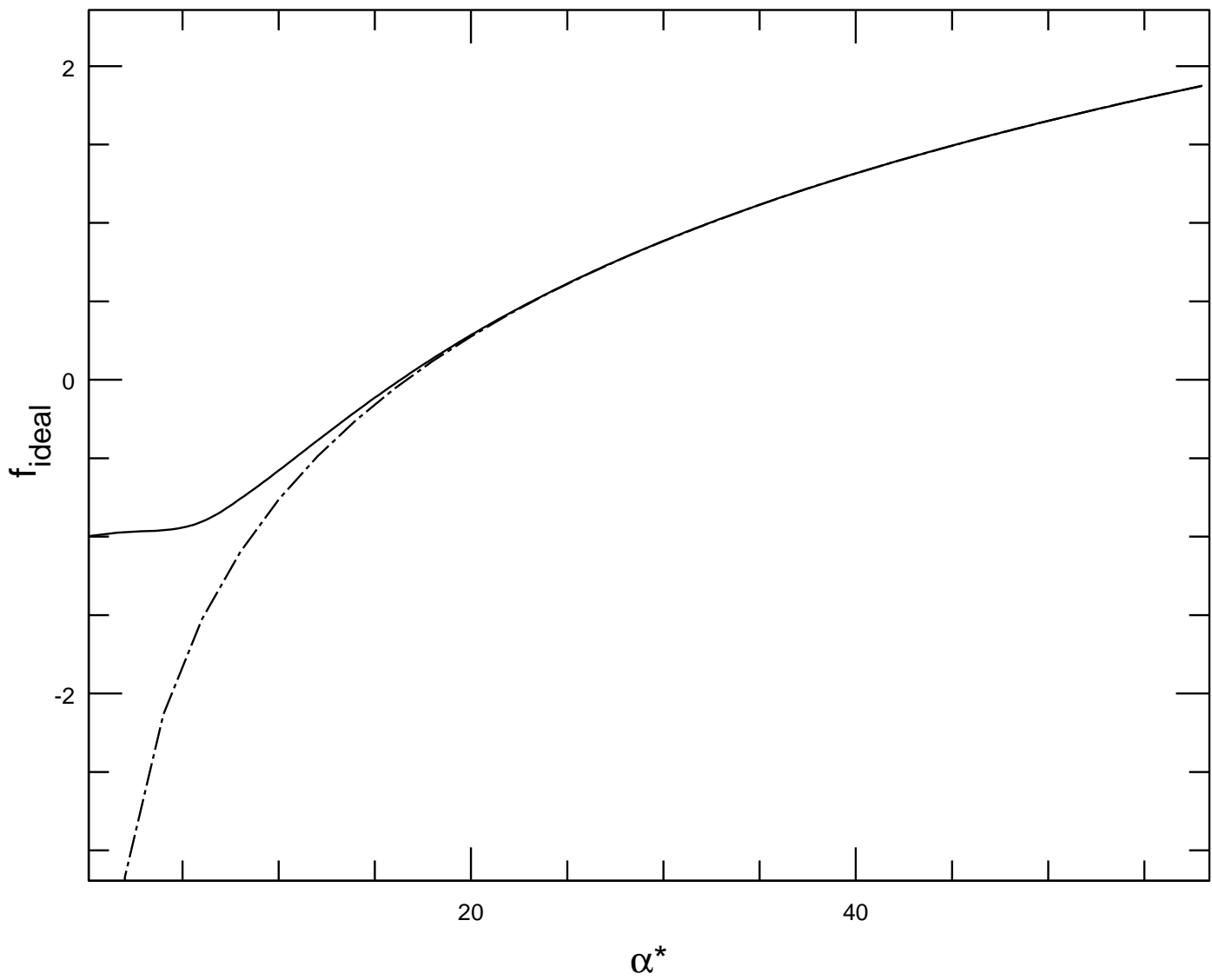
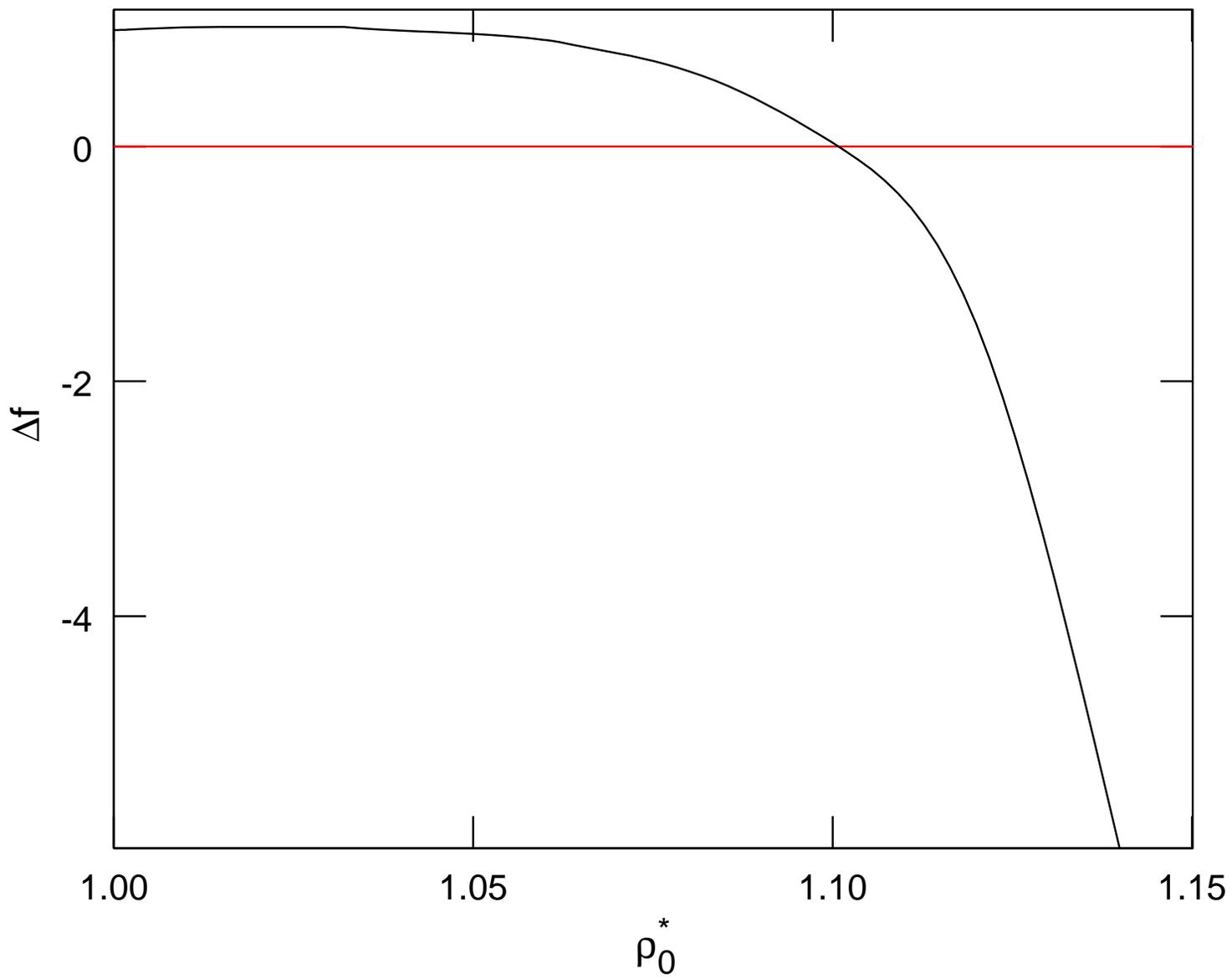
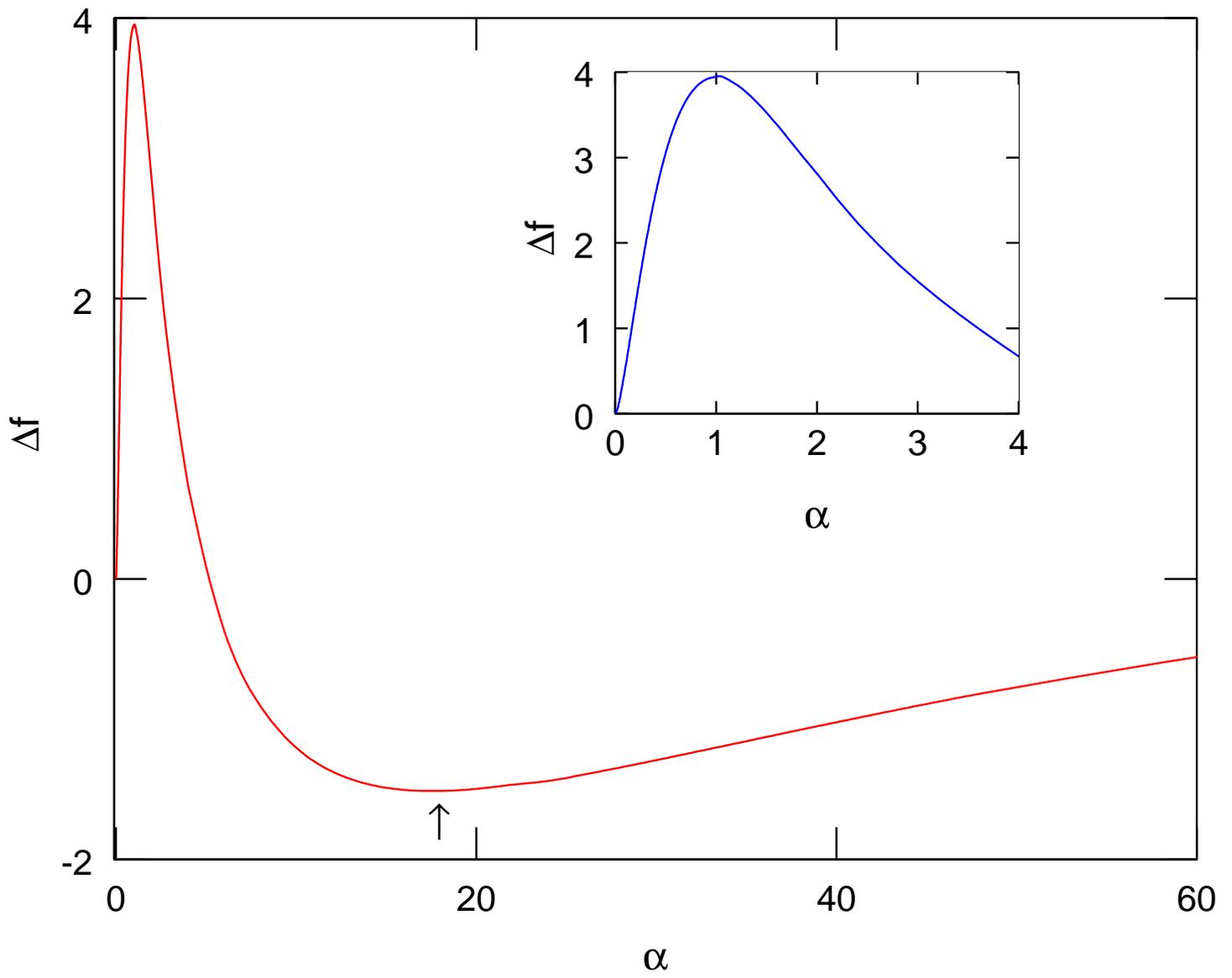
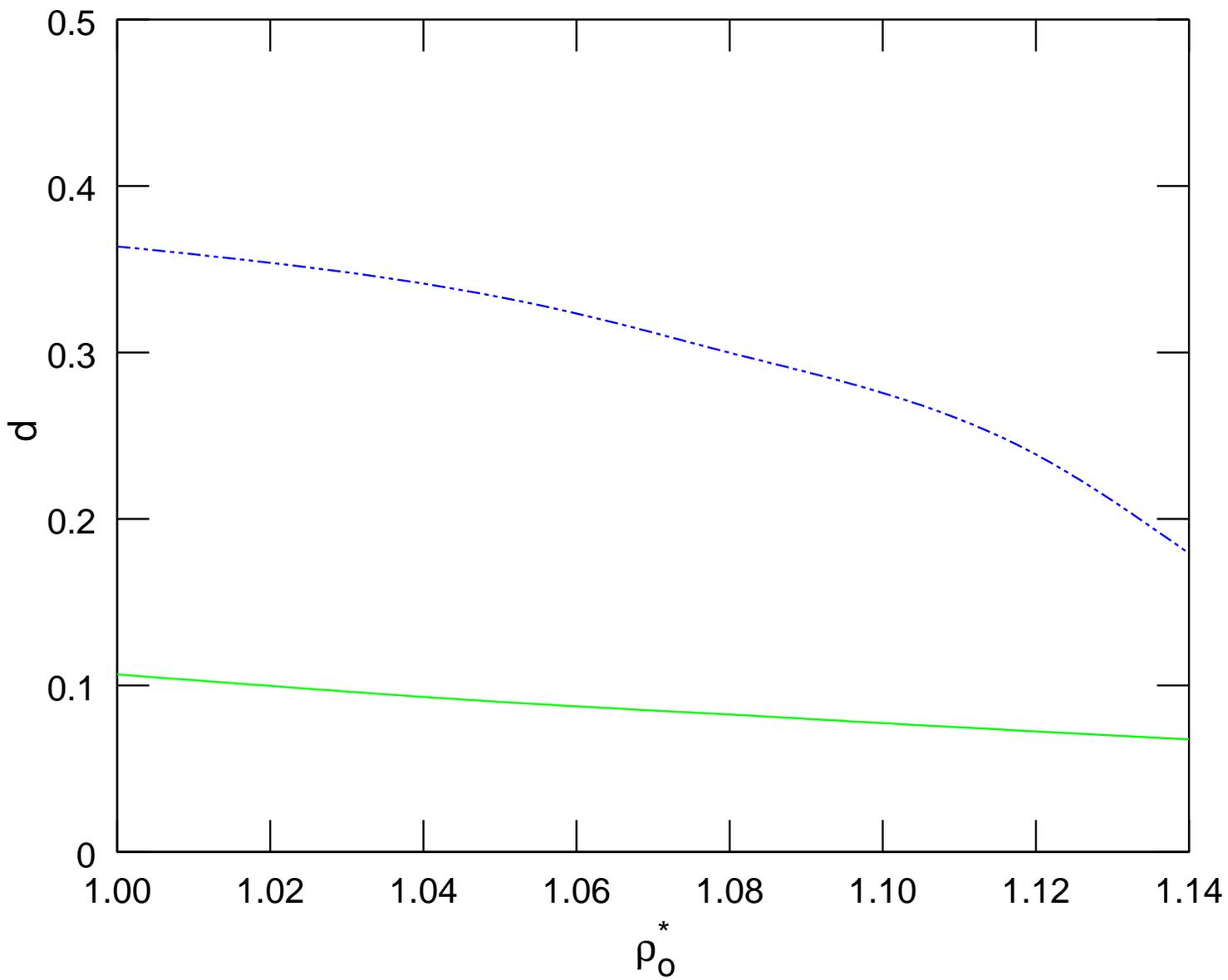
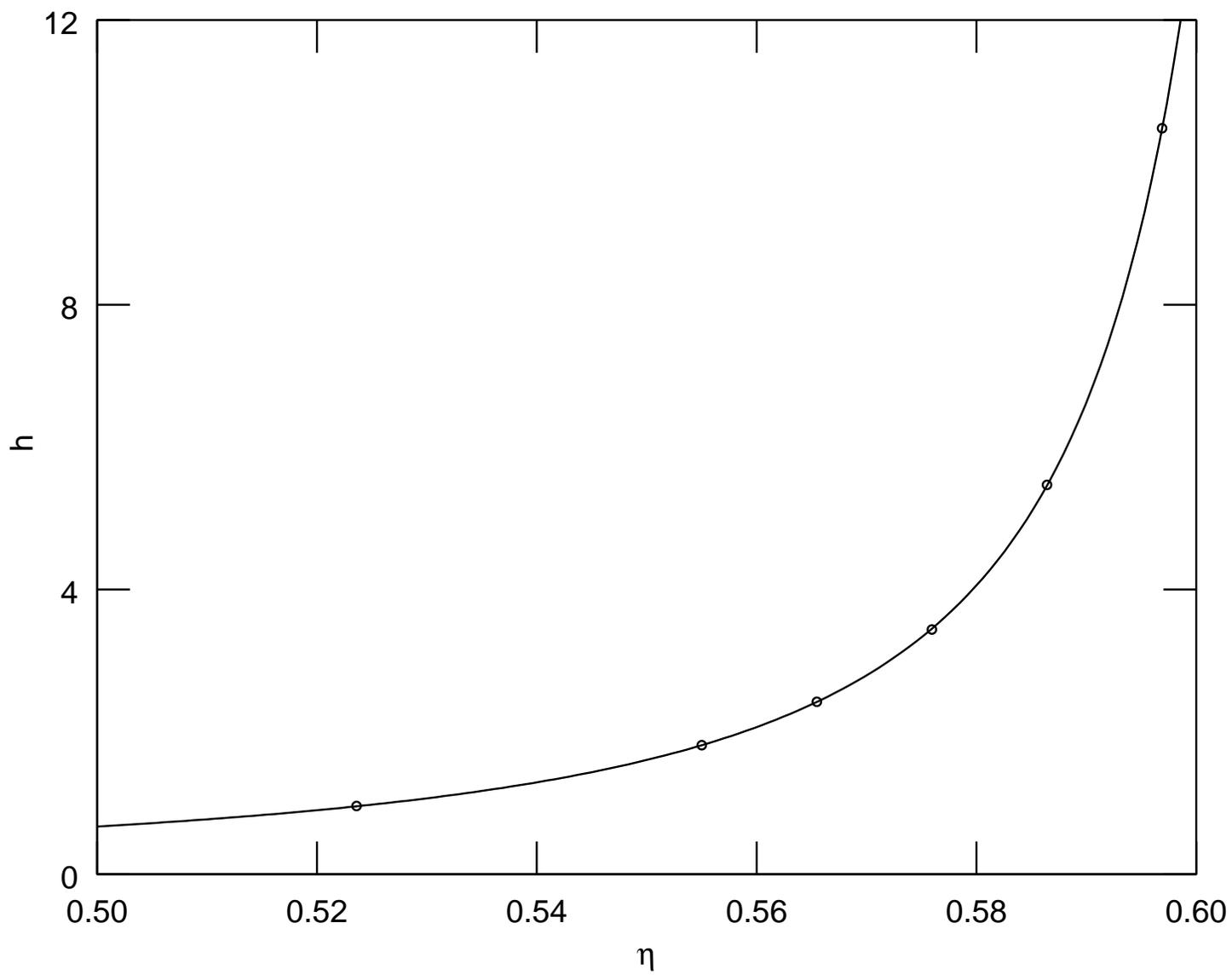
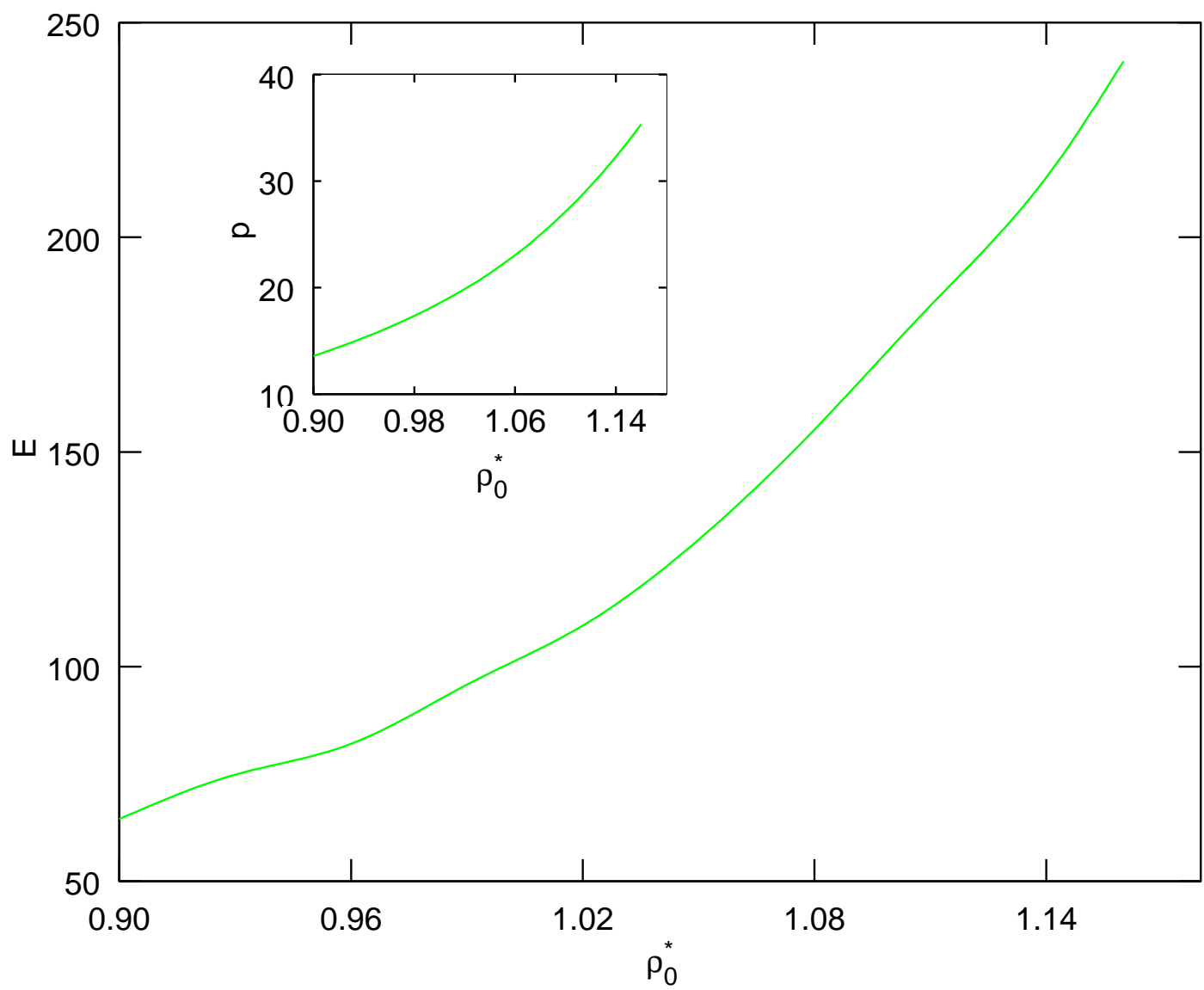
\end{document}